\documentclass[%
 aip,
 amsmath,amssymb,
reprint,%
]{revtex4-1}

\usepackage{graphicx}
\usepackage{dcolumn}
\usepackage{bm}
\usepackage[utf8]{inputenc}
\usepackage[T1]{fontenc}
\usepackage{mathptmx}
\usepackage{mathtools}
\usepackage{gensymb}

\begin{document}

\title{A versatile flexure-based $6$-axes force/torque sensor and its application to tribology} 

\author{M. Guibert}
\affiliation{Univ Lyon, Ecole Centrale de Lyon, ENISE, ENTPE, CNRS, Laboratoire de Tribologie et Dynamique des Syst\`emes LTDS, UMR 5513, F-69134, Ecully, France}
\affiliation{Now at : Univ. Lyon, Univ. Lyon1, Ens de Lyon, Ctr. de Recherche Astrophysique de Lyon, UMR5574 CNRS (France) }
\author{C. Oliver}
\author{T. Durand}
\author{T. Le Mogne}
\author{A. Le~Bot}
\author{D. Dalmas}
\author{J. Scheibert}
\author{J. Fontaine}
\affiliation{Univ Lyon, Ecole Centrale de Lyon, ENISE, ENTPE, CNRS, Laboratoire de Tribologie et Dynamique des Syst\`emes LTDS, UMR 5513, F-69134, Ecully, France}

\date{\today}

\begin{abstract}
Six-axes force/torque sensors are increasingly needed in mechanical engineering. Here, we introduce a flexure-based design for such sensors, which solves some of the drawbacks of the existing designs. In particular, it is backlash-free, it can be wirelessly monitored, it exactly enforces 90$^\circ$ angles between axes, and it enables visual inspection of the monitored system thanks to its hollow structure. We first describe the generic design, implementation and calibration procedure. We then demonstrate its capabilities through three illustration examples relevant to the field of tribology: low friction measurements under ultra-high vacuum, multi-directional friction measurements of elastomer contacts, and force/torque-based contact position monitoring.

\end{abstract}

\pacs{}

\maketitle 



\section{Introduction}
	
The simultaneous measurement of the six components of a torsor (three forces and three moments) is desired in a wide range of solid mechanics applications, including minimally invasive surgery~\cite{seibold_prototype_2005}, robotics~\cite{li_collision_2020}, fluid mechanics~\cite{aguilar_onset_2020} and tribology~\cite{hemette_friction_2018}. Existing six-axes force/torque sensors are mainly constructed using two architectures: (i) a Stewart plateform where each leg is made of a single-axis tension/compression force sensor~\cite{sorli_six-axis_1995,wang_optimal_2013,hemette_friction_2018}, or (ii) a compliant structure, the deformation of which is monitored using strain gauges~\cite{kim_design_2001,liang_design_2010}. However, both designs have drawbacks. On the one hand, the joints needed on each leg of a Stewart platform introduce backlashes and interfering forces, a problem which is more prominent when measuring smaller forces. On the other hand, the various strain gauges to be placed on a compliant structure to offer six independent outputs imply a significant number of wires connecting the sensor to the rest of the measurement chain (typically 6 degrees of freedom $\times$ 2 strain gauges $\times$ 4 wires = 48 wires). Those many wires, as well as the usual materials used for strain gauges and their glue, make it difficult to implement such 6-axes sensors in some specific environments, e.g. ultra-high vacuum.

A way of both avoiding joints and minimizing the number of wires is to measure forces and torques through the deflection of springs of known stiffness, using only 6 (one per degree of freedom) non-contact displacement sensors. Analogous solutions are often used in laboratory-made instruments for single or two-axis high-resolution force measurements (see e.g. Refs~\cite{scheibert_stress_2009,israelachvili_recent_2010,prevost_probing_2013} in the field of tribology). The use of clamped flexure beams as springs has various advantages. First, it avoids any backlash. Second, it enables fine-tuning of the beam stiffness, and thus of the sensors' range and bandwidth, through simple choices of the beam's material and dimensions. Third, such flexure beams are compatible with all types of non-contact displacement sensors (e.g. capacitive or interferometric), for which vacuum-compatible versions are commercially available.

The first scope of this article is to present the generic design, implementation and calibration procedure of a recently patented~\cite{guibert_high-precision_2016} versatile six-axes force/torque sensor, based on six flexure beams (section~\ref{sec:design}). In particular, the proposed design allows for different stiffness, resolution and frequency bandwidth among the various axes. This design, along with the calibration procedure, also ensure that all force components are measured at perfect 90$^\circ$ angles.

The second scope is to illustrate the potential of our sensor design (section~\ref{sec:results}), through three different application cases relevant to the field of tribology (the science of contact, friction and lubrication, see Ref.~\cite{vakis_modeling_2018} for a recent review). Tribology has a natural need for six-axes force/torque measurements, as simultaneous measurements of forces tangential and normal to the contact interface are required either to calculate macroscopic friction coefficients or to monitor the distribution of local stresses along the interface~\cite{scheibert_stress_2009}. It is in addition important to measure the external torques applied on the contact, because they are known to affect the local stresses, and thus the onset of sliding of the interface~\cite{scheibert_role_2010,maegawa_role_2016}.

Many fundamental investigations in the field of tribology impose additional constraints on the force/torque sensors to be used, most of which can be readily met with our design. First, clean and environment-controlled contact conditions sometimes require the contact to be placed in a ultra-high vacuum chamber~\cite{donnet_advanced_1996,martin_tribochemistry_1999}. Thanks to the few connecting wires and the materials used, the sensor can be conveniently placed within the same chamber. Second, in the case of ultra-low friction, the angle between the tangential and normal sensors needs to be precisely 90$^{\circ}$, because any misalignment can rapidly cause artefacts of the same order or larger  than the friction coefficient to be measured~\cite{burris_addressing_2009}. As we will see, such an accurate alignment, which is difficult to achieve when two separate sensors are assembled, is naturally enforced with our design. Third, to reduce the uncertainty of friction coefficient measurements~\cite{schmitz_difficulty_2005}, the raw output signals have to be of prime quality. This usually implies to place the force sensors as close as possible to the contact~\cite{gregoire_design_2021} or to align them with forces produced in the contact to minimize the torques undergone by the sensor. This requirement can rarely be fulfilled when optical observations of the contact interface are desired~\cite{sahli_evolution_2018,sahli_shear-induced_2019}. In those cases, to enable visualization, the force sensors are often shifted laterally with respect to the contact~\cite{mergel_continuum_2019,lengiewicz_finite_2020}, thus generating large torques that may increase the force measurements uncertainties. To solve this issue, the proposed sensor is hollow, which allows visual inspection of the interface through the sensor, while positioning the sensor just above the contact. Such a hollow structure can also be useful to include further electrical or thermal feed-through.

\section{Design, implementation and calibration}\label{sec:design}

Our general design consists of a six-axis flexure structure, where the shapes and sizes of the flexible beams need to be adjusted to satisfy any set of specifications about the maximum forces/torques to be measured and the expected resolutions. In practice, the design involves two stages, which can be used separately or combined in series, in particular when very different sensitivities are desired for different axes. In this section, we will illustrate the design through one of the realizations that we have performed of it so far.

	\subsection{General design}\label{sec:gendesign}
	
			The six degrees of freedom spring, which  allows an evaluation of the six components ($F_{\mathrm{x}}$, $F_{\mathrm{y}}$, $F_{\mathrm{z}}$, $M_{\mathrm{x}}$, $M_{\mathrm{y}}$, $M_{\mathrm{z}}$) of the forces/torques applied on the sensor, is made of stainless steel wires and blades. The proposed generic design of these blades makes it possible to easily change their shapes, numbers or thickness in order to change the various stiffnesses of the sensor.
			
\begin{figure}[hbt!]
	    	\centering
	    	\includegraphics[width=0.99\linewidth]{"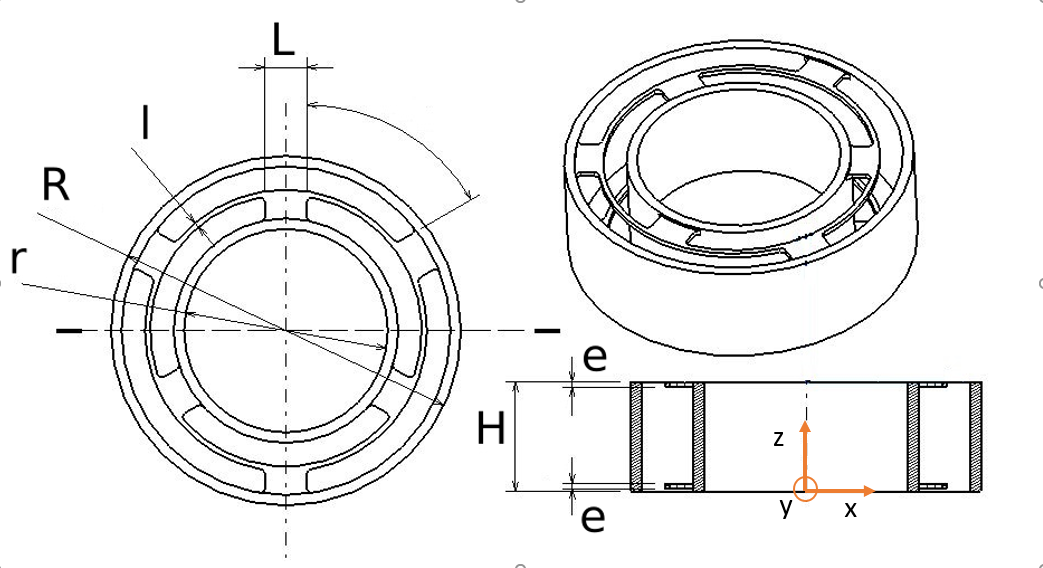"}
    		\caption{Drawing of the first stage of our sensor. The flexible parts are two identical and parallel blades connecting two concentric rigid cylinders at their top and bottom. All represented dimensions are adjustable parameters of the general design. In this realization, $R=67$~mm, $r=39$~mm, $H=21$~mm, $L=8$~mm, $l=1$~mm, $e=1$~mm.}
    		\label{fig:plans-lames}
	    \end{figure}

			For the first stage, we use a structure based on two horizontal blades positioned on two horizontal planes, as shown in  Fig.~\ref{fig:plans-lames}. All dimensions reported on the figure are those of the particular sensor we describe here, and correspond to adjustable parameters of the general design. In practice, the blades, made in 15/5PH stainless steel, are prepared by electro-erosion. This structure offers a good sensitivity in all direction, except for $M_{\mathrm{z}}$. This limitation is due to geometrical constraints of this particular realization, for which both the internal and external diameters of the sensor were imposed.

\begin{figure}[hbt!]
			    \includegraphics[width=0.99\linewidth]{"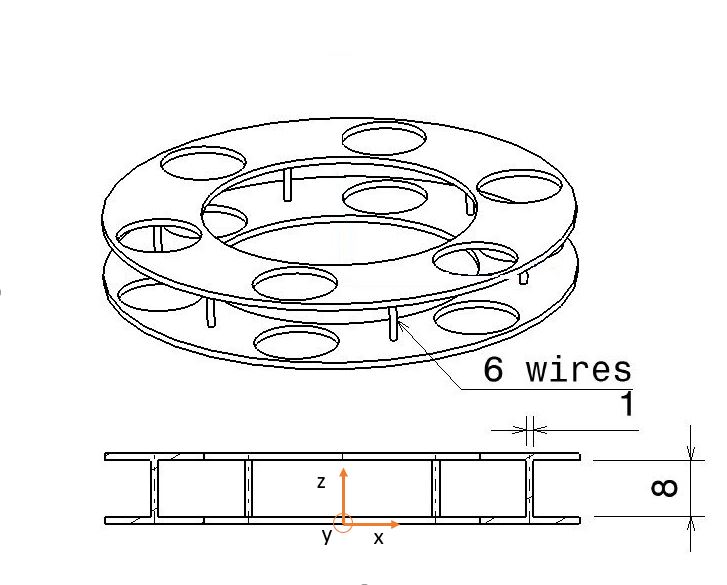"}
				\caption{Drawing of the second stage of our sensor. The flexible parts are 6 vertical wires connecting two parallel annuli. The circular holes on both annuli were added to access screws located below the stage. The represented dimensions are adjustable parameters of the general design.}
				\label{fig:plan-first-stage}
			\end{figure}

			This limitation in the $M_{\mathrm{z}}$ direction can be compensated by the use of a second stage, placed in series with the first one, and made of  $n$ vertical cylindrical rods, as shown in Fig.~\ref{fig:plan-first-stage}. In this geometry, the stiffnesses associated with $F_{\mathrm{z}}$, $M_\mathrm{x}$ and $M_\mathrm{y}$ can be considered as infinite compared to those associated with $F_{\mathrm{x}}$, $F_{\mathrm{y}}$ and $M_\mathrm{z}$. Here again, the represented dimensions show the adjusted parameters that affect the expected stiffness at first order. The whole second stage was made  in 15/5PH stainless steel, with the wires being welded to the two annuli.
			
            When the two stages are used in series, the total deflection of the sensor for a given load is the sum of the deflections of the two stages. In our realization, the first stage is more dedicated to the measurement of $F_{\mathrm{x}}$, $F_{\mathrm{y}}$, $F_{\mathrm{z}}$, $M_{\mathrm{x}}$, and $M_{\mathrm{y}}$ while the second stage adds compliance in the directions of $M_{\mathrm{z}}$, $F_{\mathrm{x}}$ and $F_{\mathrm{y}}$. The choice of using one or two stages will depend on the final application of the machine where the sensor will be installed.
            
	    	We emphasize that, for both stages, the proposed design separates the compliant parts from the stiff ones. For the first stage, the blades are clamped onto both cylinders using sliding struts. The blades are thus easily changeable, making it possible to replace them after a destruction by overload, or to adjust the resolutions and ranges of the sensor. For the second stage, the wires are not so easily changeable, but making a new stage with a different stiffness only requires different wires on identical annuli.
	    	
\subsection{Modeling-assisted conception}

Once the general design of section~\ref{sec:gendesign} is given, for each realization of a sensor, all dimensions of both stages (see Figs.~\ref{fig:plans-lames} and~\ref{fig:plan-first-stage}) need to be optimized against a precise set of specifications. In particular, the various stiffnesses and the natural frequencies are important characteristics. The former will enable an optimal matching between the maximum force to be applied to the 6-axes sensor and the measurement range of the displacement sensors, while the latter will inform about the expected bandwidth of the sensor.

\begin{figure}[hbt!]
	\centering
	\includegraphics[width=0.7\linewidth]{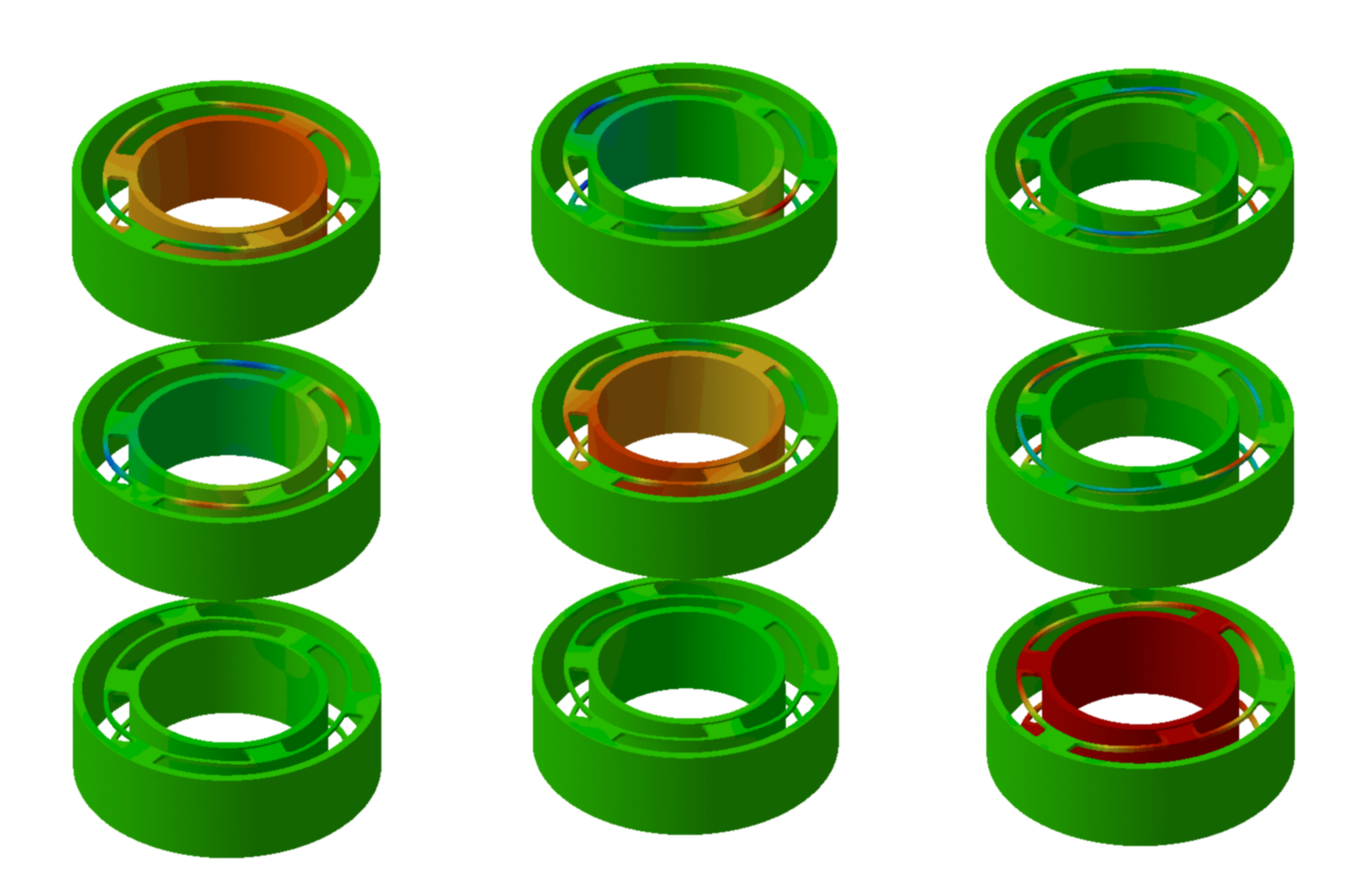}
	\caption{FEM-predicted displacement components along the $x-$, $y-$, and $z-$ axes (top, middle and bottom lines, respectively), when forces of 1\,N are applied along the $x$, $y$, and $z$ directions (left, middle and right columns, respectively). The full scale displacement (red color) is 0.2\,$\mu$m for all figures, except for the bottom right ($z$ displacement resulting from a force along the $z-$axis) where it is 5\,$\mu$m. Green indicates vanishing displacements.}
	\label{fig:forces}
\end{figure}

To access those quantities, a finite elements model (FEM) of each of the two stages has been developed using the software Catia. The model was meshed using 300~000 tetrahedral elements with a minimum size of 0.2~mm (within the thickness of the blades) and a maximum size of 3~mm (within the body of the sensor). The material behaviour was assumed to be linear elastic with a Young's modulus of 210~GPa and a Poisson's ratio of 0.31. The bottom of the outer cylinder was clamped, while the force/torques were applied to all nodes of the internal face of the inner cylinder. Figure~\ref{fig:forces} shows, for our final set of dimensions, and for the first stage of the sensor, the predicted displacements relative to the outer cylinder, when forces of magnitude 1~N are applied on the inner cylinder along all three axes. Analogously, Fig.~\ref{fig:forces} shows the same displacement components when moments of magnitude 0.01~N.m are applied to the inner cylinder around all three axes. 

\begin{figure}[hbt!]
	\centering
	\includegraphics[width=0.7\linewidth]{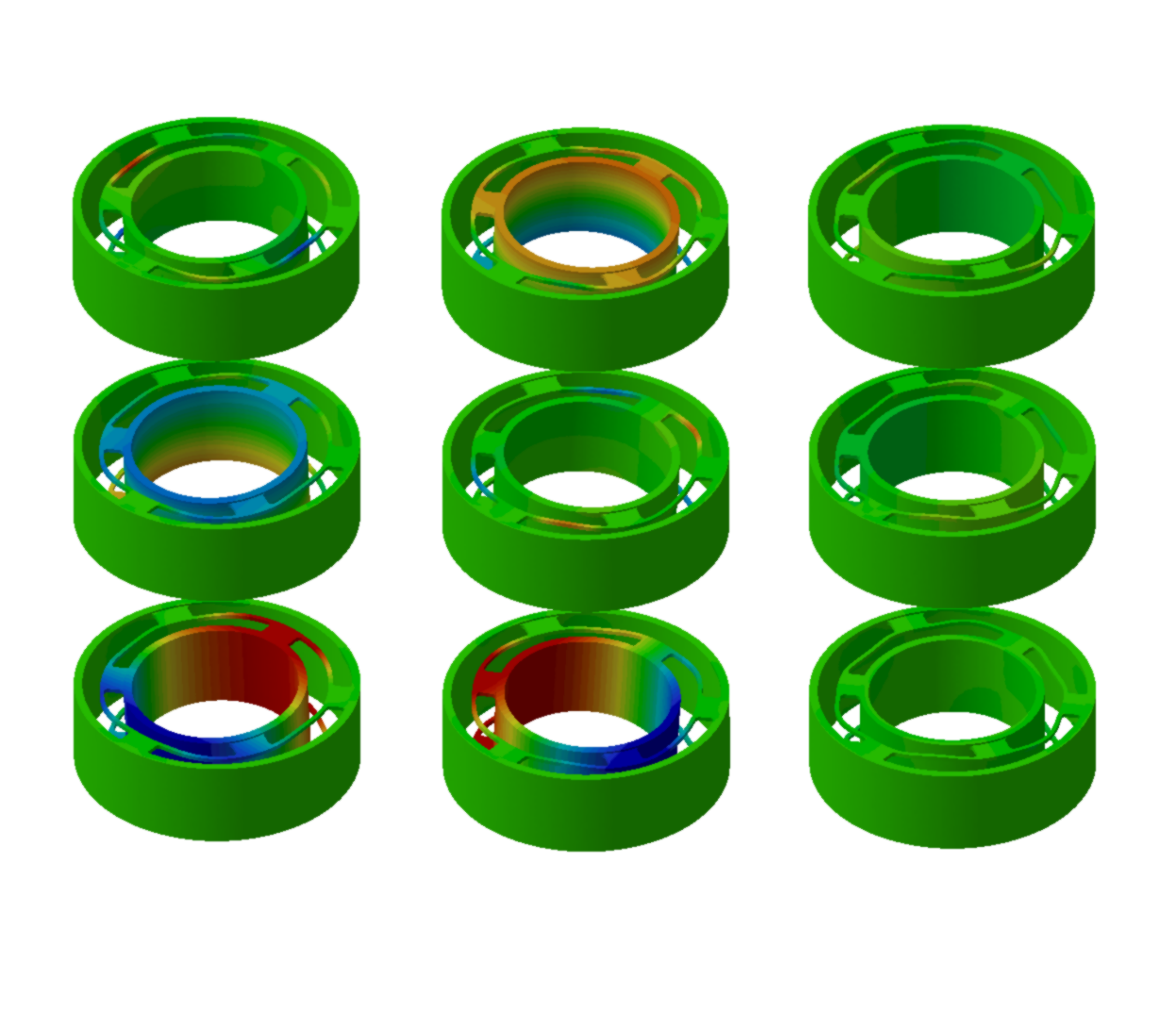}
	\caption{FEM-predicted displacement components along the $x-$, $y-$, and $z-$ axes (top, middle and bottom lines, respectively), when torques of 0.01\,N.m are applied around the $x$, $y$, and $z$ directions (left, middle and right columns, respectively). The full scale displacement (from red to blue color) range from -0.2\,$\mu$m to 0.2\,$\mu$m for all figures. Green indicates vanishing displacements.}
	\label{fig:moments}
\end{figure}

As a final result of those FEM calculations, the various translation stiffnesses of the first stage are 5.10$^6$\,N/m for the $x$ and $y$ axis, and 2.10$^5$\,N/m along the $z$ direction. The high values of stiffness along $x$ and $y$ directions are lowered using the second stage while the stiffness along the Z direction is not modified. The rotational stiffnesses, for the first stage, are 0.3\,rad/(N.m) for the $x$ and $y$ axis and 33\,rad/(N.m) around the $z$ directions, respectively. Here again, the high value of the stiffness along the $z$ direction is lowered by the second stage, while the ones along the $x$ and $y$ directions are not changed. For all these forces and moments cases, maximum Von Mises strain obtained is around 5\,MPa. These values can be used to obtain the maximum displacement and strain resulting from a 10~N force along the $x$, $y$ and $z$ directions. Those are about 70\,$\mu$m and 70\,MPa respectively.

We also used our FEM model to perform a modal analysis of each of the two stages. All eigenfrequencies were found larger than about 120~Hz, so that the behaviour of the stages can be considered as quasistatic below this typical frequency. Note that, due to the additional masses attached to the stages in a real measurement environment, the measurement bandwidth is in practice smaller than the quasistatic range of the individual stages. In our case (see Figs.~\ref{fig:capteur3dsketch} and~\ref{fig:capteur3d}), this practical bandwidth was estimated experimentally to be about 80\,Hz.

\subsection{Displacement measurements}

To extract the six components of forces/torques applied on the sensor, we now need to measure accurately the relative displacements between its rigid extremities (cylinders and/or annuli, depending whether one or both stages are used). Indeed, due to the flexible elements in each stage, any force applied on the sensor will lead to a small relative displacement. To avoid any parasitical force, those displacement are measured using non-contact sensors. In the case of our realization, based on the FEM-estimated stiffnesses and bandwidth of the 6-axes sensor, and on its desired resolution, we had the following constrains for the choice of the displacement sensors: a resolution of a few nanometers (enabling a force resolution bellow one milli-Newton), a measurement range (corresponding to a maximum relative displacement of the sensor's extremities) of one hundred micrometers and a bandwidth of a few hundred Hertz (larger than the sensor's bandwidth).
These characteristics led us to consider two different technologies, Michelson interferometric sensors or capacitive sensors, among which we chose the latter for price reasons.

\begin{figure}[hbt!]
	\centering
	\includegraphics[width=1.0\linewidth]{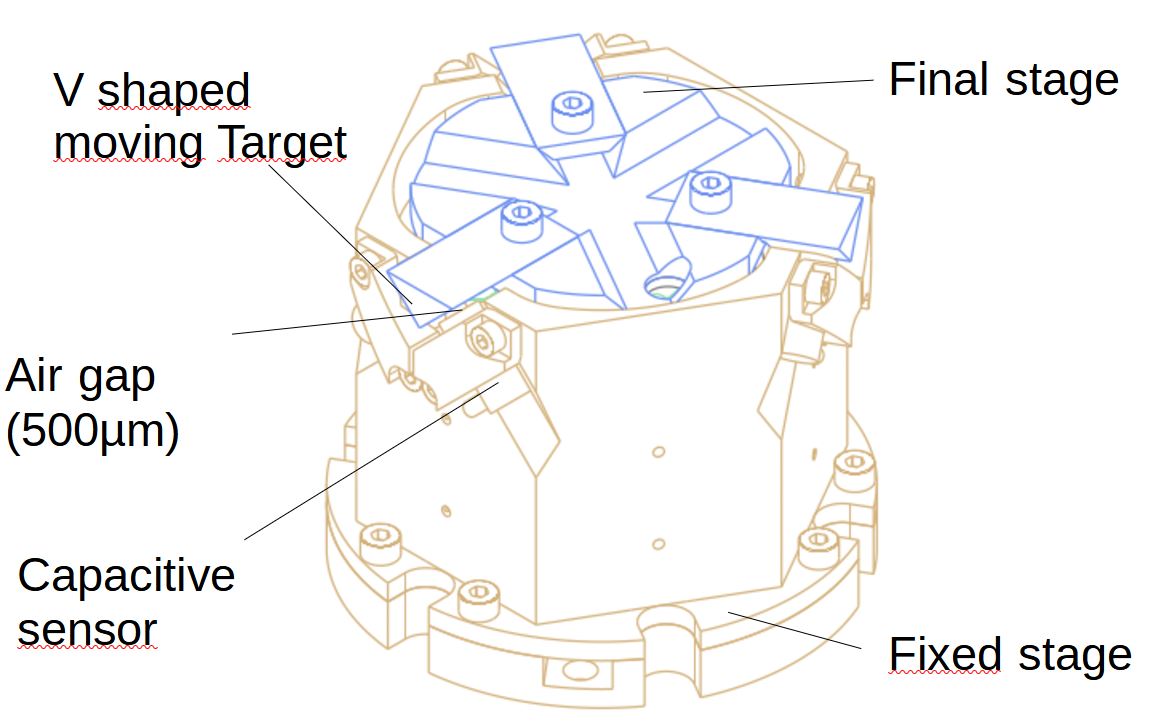}
	\caption{Schematic representation of the 6-axes sensor in its final environment: a casing ( brown) supporting the six capacitive displacement sensors, and three triangular arms (blue), the faces of which serve as targets for those sensors.}
	\label{fig:capteur3dsketch}
\end{figure}

\begin{figure}[hbt!]
	\centering
	\includegraphics[width=1.0\linewidth]{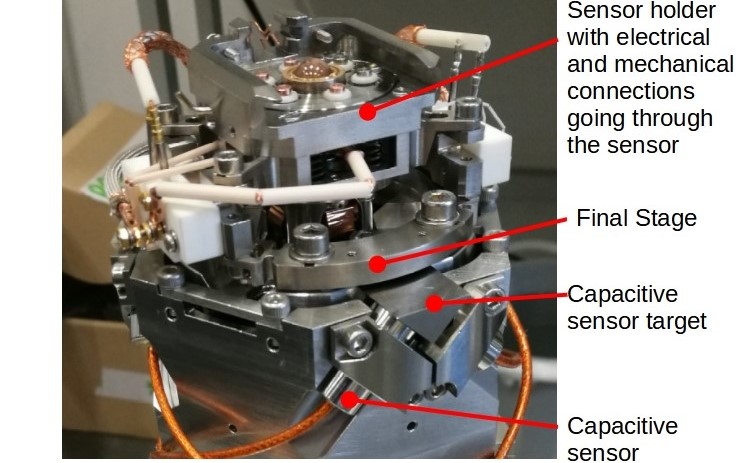}
	\caption{Picture of the realized 6-axes sensor. The dock for sample holder (not shown in Fig.~\ref{fig:capteur3dsketch}) with electrical connections going through the central hole of the sensor can be seen on its top.}
	\label{fig:capteur3d}
\end{figure}

Figure~\ref{fig:capteur3dsketch} is a sketch of the casing in which the realized 6-axes sensor has been placed, in order to hold the six displacement sensors. The outer cylinder of the first stage is fixed in the brown casing, while the top annulus of the second stage, which is free, is equipped with three arms (blue part) having a triangular shape and distributed at 120$^{\circ}$ of each other. These arms are the target used by the capacitive sensors for their displacement measurements. Thus, the displacement measurement is constructed as a Stewart platform, where six capacitive sensors (Fogale Nanotech MCC with conditioner MC900) are positioned by pairs (each sensor points toward one of the two bottom faces of one triangle), oriented at 45$^{\circ}$ with respect to the $z-$axis (vertical axis in Fig.~\ref{fig:capteur3dsketch}). These displacement sensors have a resolution better than 1\,nm  at a frequency of 1\,Hz. Figure~\ref{fig:capteur3d} is a picture of the actual sensor, on which we can see, on top of the final stage holding the triangular targets, the sample holder and the electrical connections passing through the hollow structure of the sensor.
	
\subsection{Force calibration}
Instead of calculating the applied forces as a product of displacement and stiffness, and to avoid propagation of any error on the stiffness values of the flexible stages, we realized an \textit{in~situ} calibration in which the outputs of the six displacement sensors are directly related to known applied forces. Because the values of the displacements (of the order of 100\,$\mu$m) are much smaller than the macrometric size of the sensor (about 50\,mm), all strains and rotation angles remain small, which allows us to complete the calculations considering a linear system. The applied procedure is described in Ref.~\onlinecite{barbato_7_nodate}, and does not require any prior calibration of the displacement sensors, since only their linearity is mandatory. It consists in the application of a number of elementary forces on the sensor, each with known application point, direction and magnitude.

\begin{figure}[hbt!]
	\centering
	\includegraphics[width=0.99\linewidth]{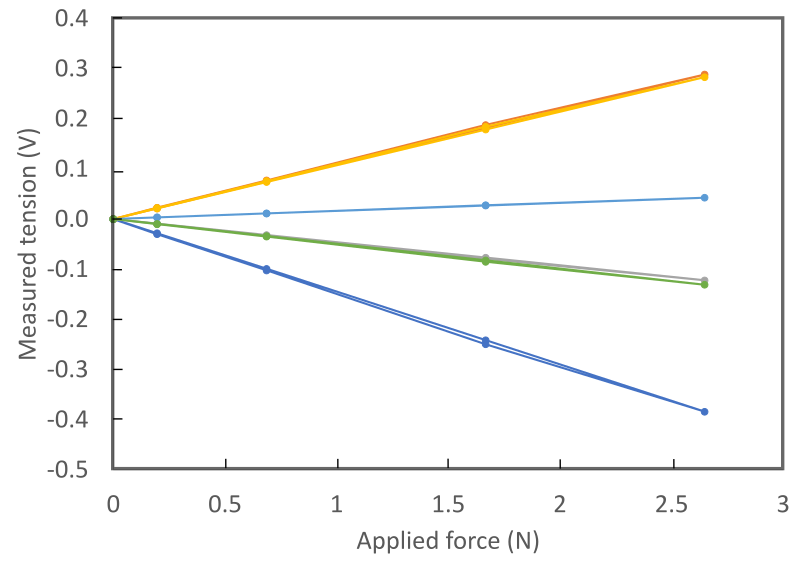}
	\caption{Outputs of the 6 displacement sensors (one color per sensor; note that the orange and yellow, as well as the purple and green curves, are hardly distinguishable), as a function of the magnitude of the force applied along the $x$-axis, on a single particular location. The figure shows loading and unloading curves.}
	\label{fig:linearite_capteur}
\end{figure}

To reduce the uncertainties on the applied forces, a dedicated mechanical part has been developed, which enables application of dead weights at specific positions to stimulate the $F_{\mathrm{z}}$, $M_\mathrm{x}$ and $M_\mathrm{y}$ components. For the horizontal axes, we use pulleys and strings for the calibration of $F_{\mathrm{x}}$, $F_\mathrm{y}$ and $M_\mathrm{z}$. For each axis ($x$, $y$ or $z$), forces are applied at four different locations. At each location, ten different magnitudes are tested, covering the full range of the sensor, in an increasing-then-decreasing manner. Doing so, one could detect any unwanted non-linearity or hysteresis on the sensor. Typical measurements obtained during the calibration campaign, for a single location, are shown in Fig.~\ref{fig:linearite_capteur}.

The full calibration of the sensor led to 120 different measurements, from which we define two 6$\times$120 matrices, $F_{cal}$ and $U_{cal}$, containing respectively the six forces/torque components of the applied stimuli and the six outputs of the displacement sensors. The desired exploitation matrix is thus the 6$\times$6 matrix $A$, such that, formally: $F_{cal}=A.U_{cal}$. Solving for $A$, in such an over-determined system, is performed in the least-square sense and gives:

\begin{widetext}
\begin{equation}
    A=
    \begin{pmatrix*}[r]
           5.394  &   -19.524  &	13.264  &   14.344  &     5.609  &   -18.427 \\
         -18.187  &    -4.980  &    13.793  &  -14.391  &    19.218  &     4.555 \\
          -3.522  &    -3.802  &    -3.279  &   -3.443  &    -3.490  &    -3.621 \\
        2607.080  &  1460.757  & -1463.105  & 1293.960  & -2677.257  & -1129.515\\
          -4.318  & -2363.428  &  2233.907  & 2334.530  &    46.965  & -2229.395\\
         158.344  &  -128.460  &   126.039  &  -86.929  &  -160.759  &    89.937\\
    \end{pmatrix*}
\label{matrice_tec}
\end{equation}
\end{widetext}

when units used during calibration are Newton, Newton$\cdot$millimeter and Volts. For any new measurement, the six estimated force/torque components are the entries of the $6\times 1$ vector $F$ calculated as
\begin{equation}
    F=A\cdot U,\label{eq:FPGA}
\end{equation}
where the entries of the $6\times 1$ vector $U$ are the six outputs of the six displacement sensors.

The first (resp. second and third) line of $A$ contains the coefficients to be applied to each displacement output to obtain the force applied to the sensor along the $x$ (resp. $y$ and $z$) direction. Analogously, the three last lines contain the coefficients needed to obtain the torques applied on the sensor. Note that all coefficients in the third line ($F_z$) are negative, because all the distances between displacement sensors and their respective targets decrease when $F_z$ increases. More generally, the sign of each coefficient indicates the direction of motion of the target of each displacement sensor when the corresponding force/torque component is applied. The quasi-absence of vanishing coefficients shows that each force measurement depends non-negligibly on all displacement sensors. A practical consequence of this property is that checking the calibration of the sensor in a single direction is actually sufficient to check it in all directions. Such a possibility is especially useful when accessing the sensor is difficult, for instance in ultra-high vacuum conditions. A new full calibration is needed only when a rapid check on a single axis (which can easily be repeated frequently) yields an unexpected results.

\subsection{FPGA implementation}\label{sec:FPGA}
To enable real time evaluation of the forces applied on the sensors, the matrix product of Eq.~\ref{eq:FPGA} is directly implemented on a National Instruments C-RIO 9237 FPGA (Field-Programmable Gate Array) target. It converts the six measured voltages through two NI9239 delta-sigma 24-bits anti-aliased analog-to-digital converters, into the six components of the force/torque vector. With such a hardware implementation, no calculation is done on a PC hardware, which ensures a perfectly controlled time delay between the voltage measurements and the force vector calculation. In addition, it enables dynamic measurements, and offers the possibility to perform a real-time closed-loop control of on the applied force on the tribometer. Using anti-aliased converters on the FPGA allows us to implement numerical low pass filter for the force measurements. This way, each sensor output can be ascribed a different cut-off frequency, allowing to extract both the static and dynamic responses of the sensor. Finally, our FPGA implementation incorporates the possibility to express the six component force vector in any projection base. One may therefore express the forces in a base that is not aligned anymore with the sensor, but with more physically interesting axes. For instance, in tribology applications, we often express the forces in a frame in which the normal load is applied along the $z$ axis, even though the sensor and tribometer may not be perfectly aligned. This in particular applies for the results of section~\ref{sec:results}.

The FPGA implementation is done using fixed point numbers. To do so, we chose to express the matrix calibrations in N/V and N.mm/V. The obtained values, shown in Eq.~\ref{matrice_tec}, ensure values in the range 0 to 1000. Calculations are done using 32 bits fixed-point numbers having 16 digits for their integer part, and 16 for their fractional part. Since the measured voltages are in the range $\pm 10$ Volts, the maximum value of the coefficients of the matrix product $A\cdot U$ will remain smaller than $\ 6\times10\times1000<2^{16}$. Precision is therefore preserved, and no overflow event can occur, during all calculations. 

\section{Results and discussion}\label{sec:results}

In this section, we will illustrate the capabilities of the type of sensor described in section~\ref{sec:design}, through various tribology experiments. To this end, two different realizations of our force/torque sensor have been installed on two different tribometers. The first sensor is the exact one that was taken as an example throughout section~\ref{sec:design}, mounted in a ultra-high vacuum tribometer enabling experiments under gas pressure down to 1$0^{-9}$~mbar, and with a temperature ranging from -150 to 600$^{\circ}$C. It was used to run friction tests on Diamond-Like Carbon coatings under Ultra High Vacuum (section~\ref{sec:DLC}). The second sensor (see its dimensions in Appendix) was mounted on a multi-axis contact mechanics test rig enabling both dynamic excitation and in situ visualization of soft material contact interfaces. It was used both to investigate the incipient tangential loading of an elastomer contact in various directions (section~\ref{sec:PDMS}), and to locate the moving contact point of a steel ball on a glass substrate (section~\ref{sec:localization}).

\subsection{Low friction under ultra-high vacuum}\label{sec:DLC}

Ball-on-flat linear reciprocating sliding experiments are conducted under ultra-high vacuum ($<10^{-6}$\,Pa), the sensor and its dock for receiving sample holder (Fig.~\ref{fig:capteur3d}) being mounted inside a vacuum chamber. A counter-body is attached upside-down to an upper arm allowing $x$, $y$, $z$ and $\theta_z$ movements. These movements allow the relative positioning of the two samples ($x$, $y$ and $\theta_z$), the application of a controlled load ($z$) and a linear reciprocating motion ($x$, $y$). The chamber and the vacuum components have been manufactured by PREVAC (\url{https://www.prevac.eu/en/news,10/110,tribometer-chamber.html}). The experiments are conducted between a Diamond-Like Carbon (DLC) coated silicon flat (lower specimen) and a hemispherical pin (52100 steel) with a radius of curvature of 8\,mm. The DLC is an hydrogenated amorphous carbon (a-C:H) with about 42~at.\% of hydrogen. This combination of materials is known to lead to very low friction coefficients under ultra-high vacuum, with values of the coefficient of friction lower than 0.01~\cite{fontaine_JET222}. The friction experiment is conducted under a normal load of 3\,N applied through the $z$-motion of the upper specimen, and using a feedback loop in order to keep the applied load as close as possible to this value. The upper specimen is moving along the $x$-axis with 2\,mm linear reciprocating motion at a sliding speed of 0.5\,mm/s.

\begin{figure}[hbt!]
	\centering
	\includegraphics[width=0.95\linewidth]{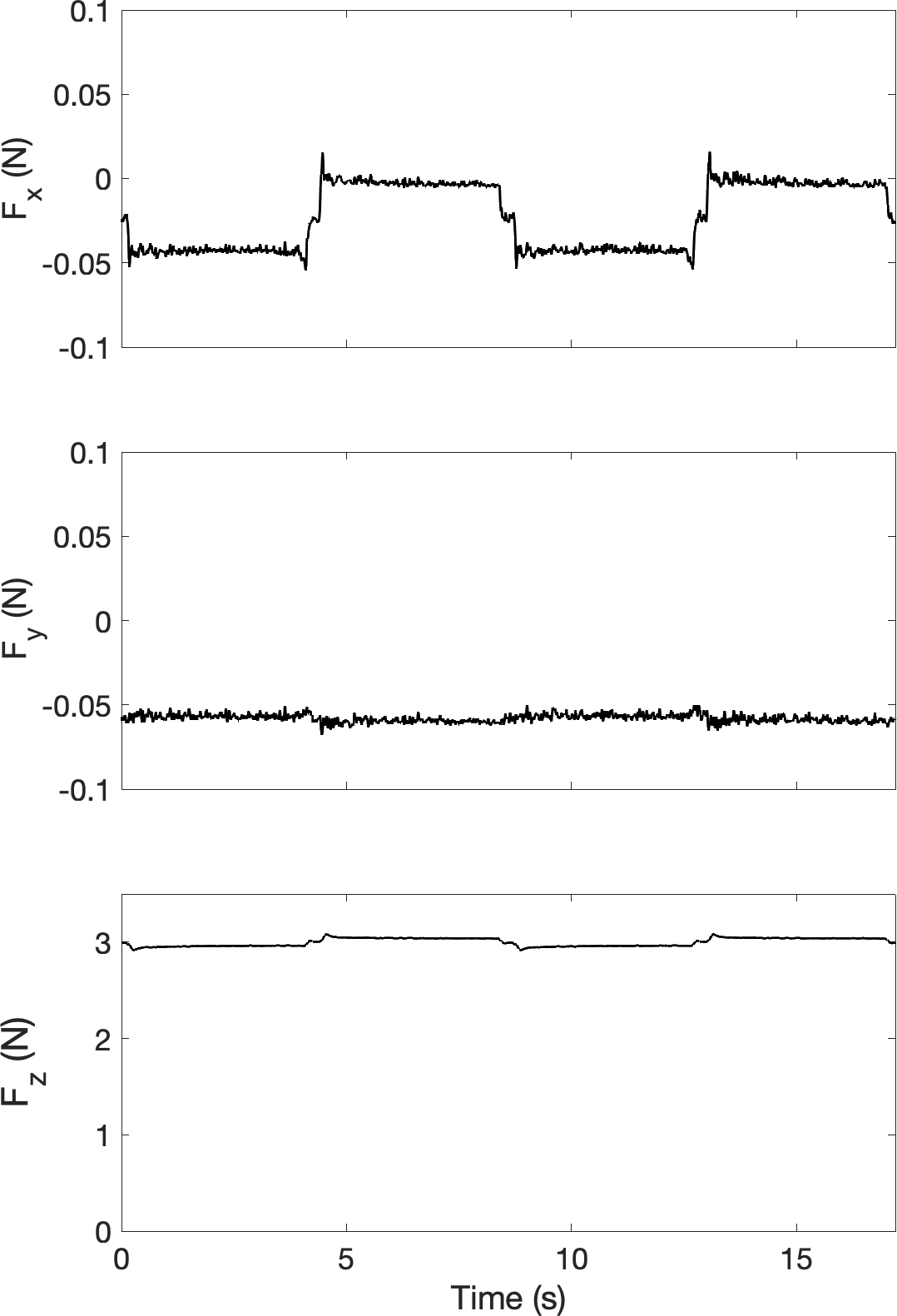}
	\caption{Time evolution of the raw measured forces, $F_x$, $F_y$ and $F_z$, during reciprocating passes 189 to 192. Note the different scale for $F_z$.}
	\label{fig:UHV1Raw}
\end{figure}

The goal of such tribological experiments is to study the evolution of the coefficient of friction with the number of reciprocating sliding cycles, in an attempt to understand the effect of morphological, mechanical and/or chemical changes occurring on the rubbed surfaces. Accurate measurements of the coefficient of friction are thus paramount.

Figure~\ref{fig:UHV1Raw} presents the evolution of raw forces, $F_x$, $F_y$ and $F_z$, during reciprocating passes 189 to 192 (a sliding cycle includes two passes, one forward and one reverse motion). The value of the force $F_x$ is alternating as expected between forward and reverse motion, although it is not centered around 0. The force $F_y$ is almost constant, as one would expect for a motion along $x$ direction, but it is slightly different from zero. The normal force $F_z$ is essentially constant, with slight variations around the 3\,N set point. These small variations can be attributed to the response delay of the feedback loop. There are also variations of the forces at the beginning and end of each sliding pass (forward or reverse), related to the time required to change the direction of motion. In the following, to disregard those transients, the data are truncated by 5\% both at the beginning and at the end of each pass.

\begin{figure}[hbt!]
	\centering
	\includegraphics[width=0.95\linewidth]{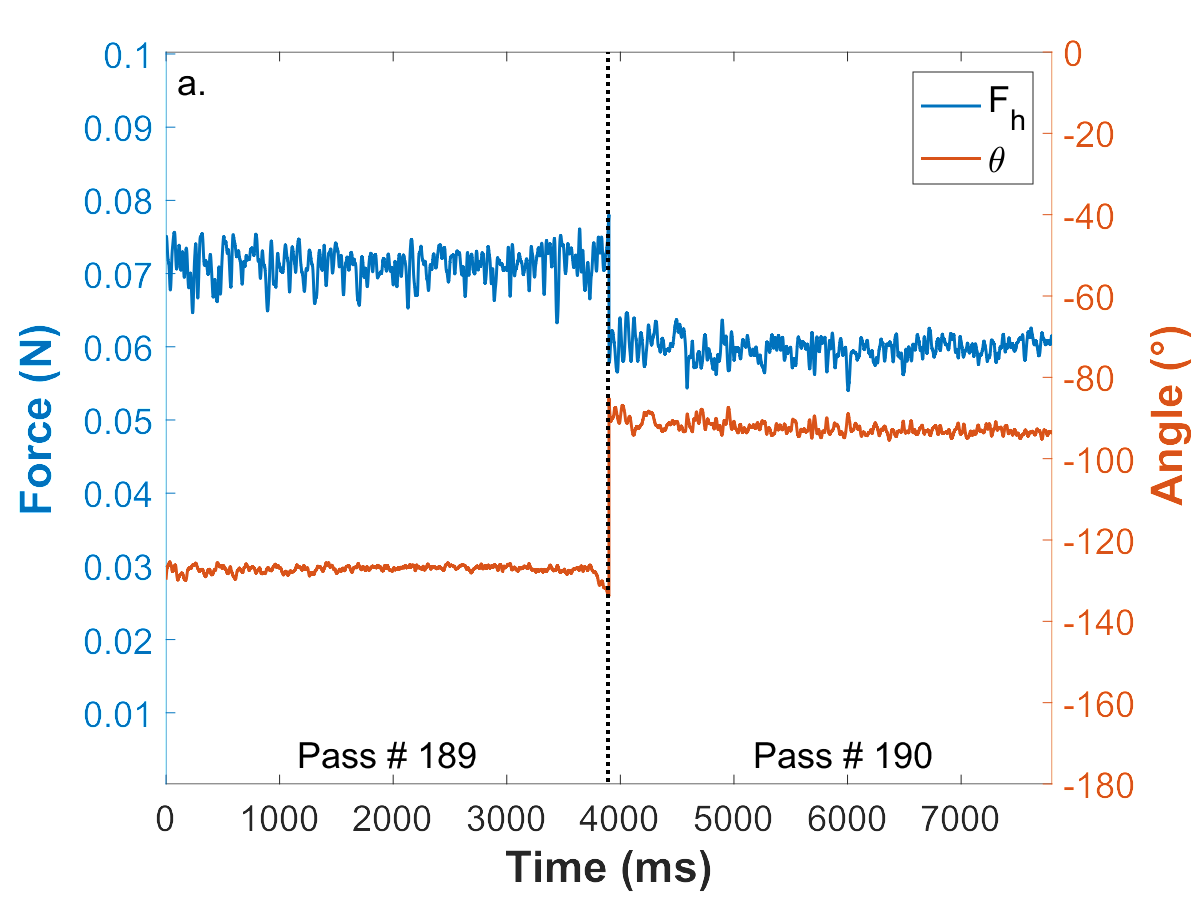}
	\includegraphics[width=0.95\linewidth]{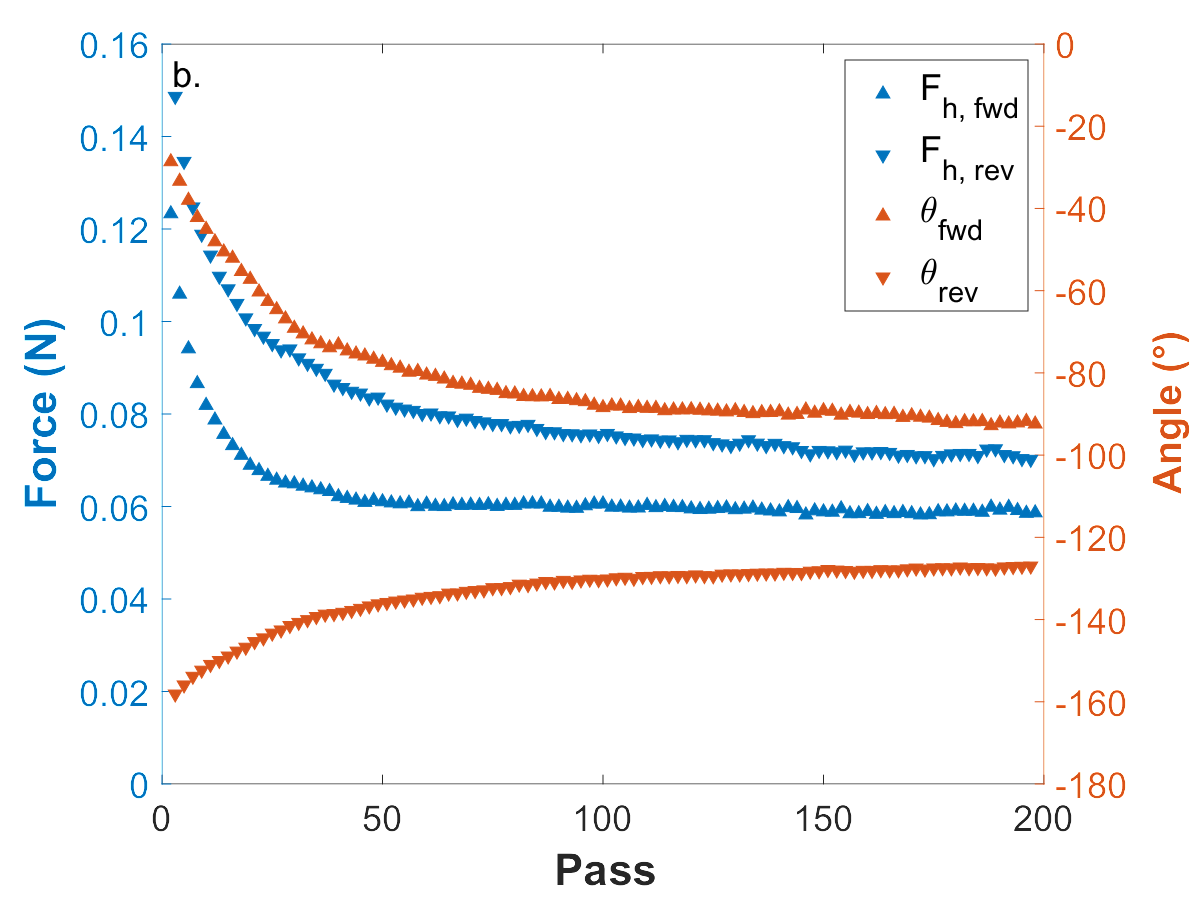}
	\caption{Evolution of the norm $F_h$ (blue) and angle $\theta$ (red) of the horizontal force $\Vec{F_h}$, (a) vs. time during reciprocating passes 189 \& 190 and (b) vs. number of sliding passes for forward ($\blacktriangle$) and reverse ($\blacktriangledown$) motion.}
	\label{fig:UHV2Raw}
\end{figure}

\begin{figure}[hbt!]
	\centering
	\includegraphics[width=0.95\linewidth]{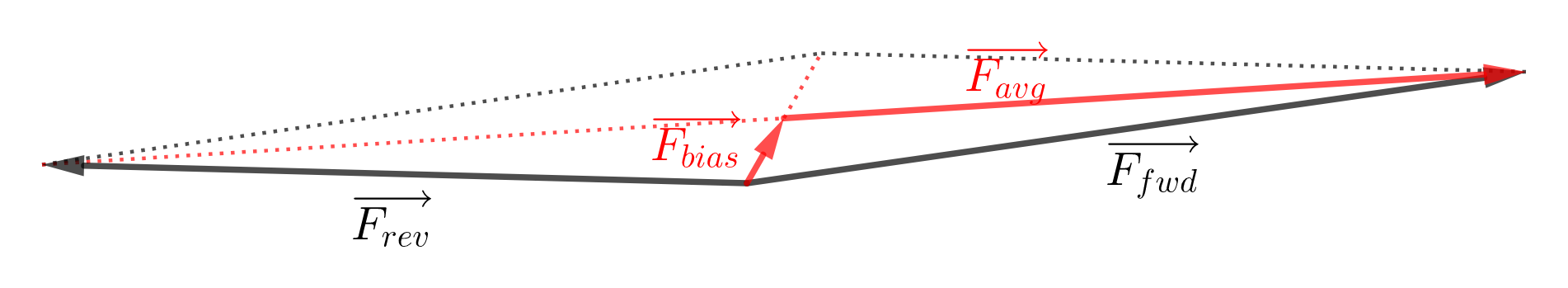}
	\caption{Definition of the average horizontal force $\Vec{F}_{avg}$ and the bias horizontal force $\Vec{F}_{bias}$ as a function of the raw horizontal forces at the same positions at forward and reverse motion $\Vec{F}_{fwd}$ and $\Vec{F}_{rev}$, according to equations~\ref{eq:Favg} \& \ref{eq:Fbias}.}
	\label{fig:UHV3Vectors}
\end{figure}

From the measured values of $F_x$ and $F_y$, the horizontal (in a plane orthogonal to $z$) force vector $\overrightarrow{F_h}$ is defined with these coordinates. The norm $F_h$ and angle with respect to the $x$-axis, $\theta$, of this horizontal force can be then easily computed. The time evolution of both values, plotted on Fig.~\ref{fig:UHV2Raw}.a, clearly shows that there is a constant bias of the norm between forward and reverse direction, and the change in angle is much less than the 180\,$\degree$ expected for a reciprocating motion. This can also be seen as a function of sliding cycles on Fig.~\ref{fig:UHV2Raw}.b: the change in angle $\theta$ between forward and reverse motion seems to decrease with the norm, suggesting that the measured force $\overrightarrow{F_h}$ is not the actual tangential force. There seems to be a contribution of a bias force that should be subtracted to get the real tangential force. As shown on Fig.~\ref{fig:UHV3Vectors}, we can use the forward and reverse horizontal forces $\overrightarrow{F_{fwd}}$ and $\overrightarrow{F_{rev}}$ of two consecutive sliding passes to estimate the average tangential force $\overrightarrow{F_{avg}}$ and the residual force $\overrightarrow{F_{bias}}$, by using the following equations:
\begin{equation}
    \overrightarrow{F_{avg}}=\frac{\overrightarrow{F_{fwd}}-\overrightarrow{F_{rev}}}{2}, \label{eq:Favg}
\end{equation}
\begin{equation}
    \overrightarrow{F_{bias}}=\frac{\overrightarrow{F_{fwd}}+\overrightarrow{F_{rev}}}{2}. \label{eq:Fbias}
\end{equation}

\begin{figure}[hbt!]
	\centering
	\includegraphics[width=0.95\linewidth]{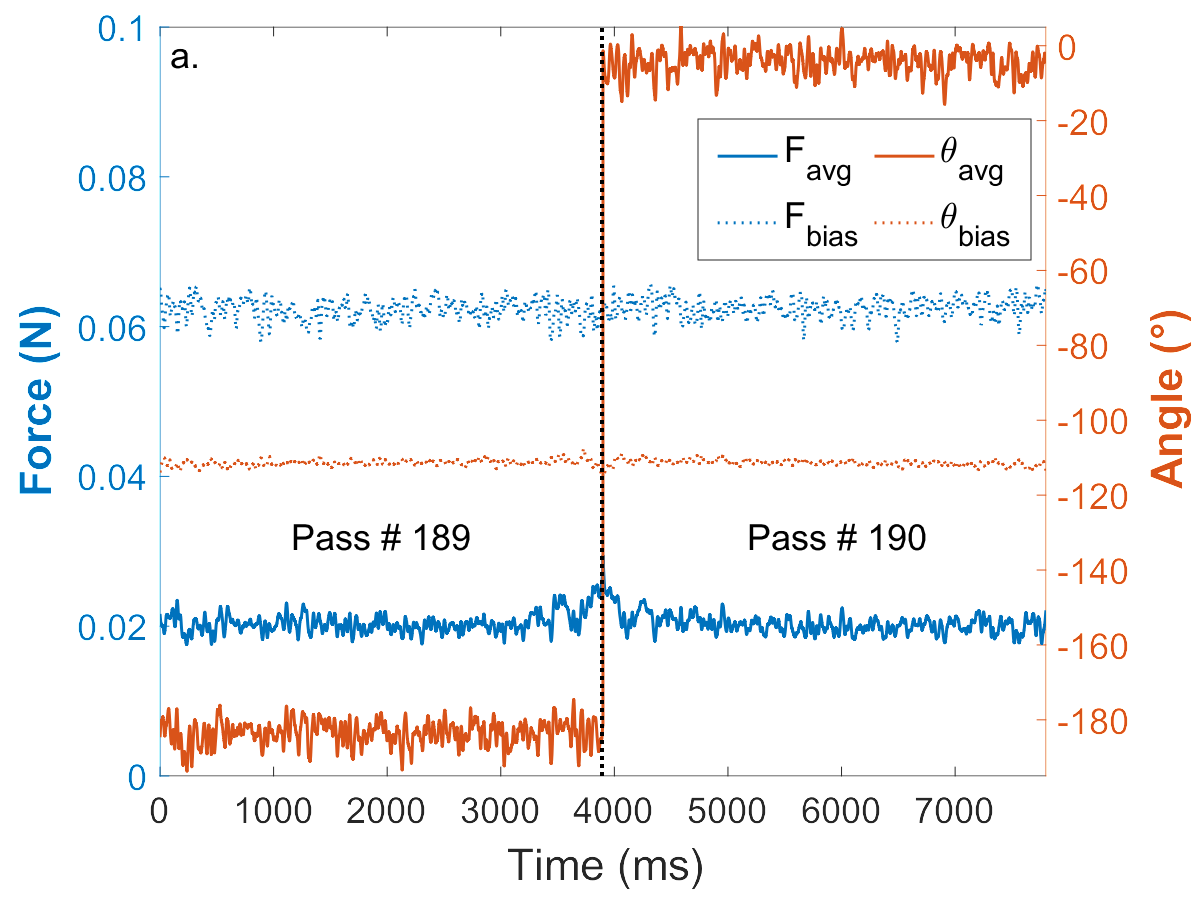}
	\includegraphics[width=0.95\linewidth]{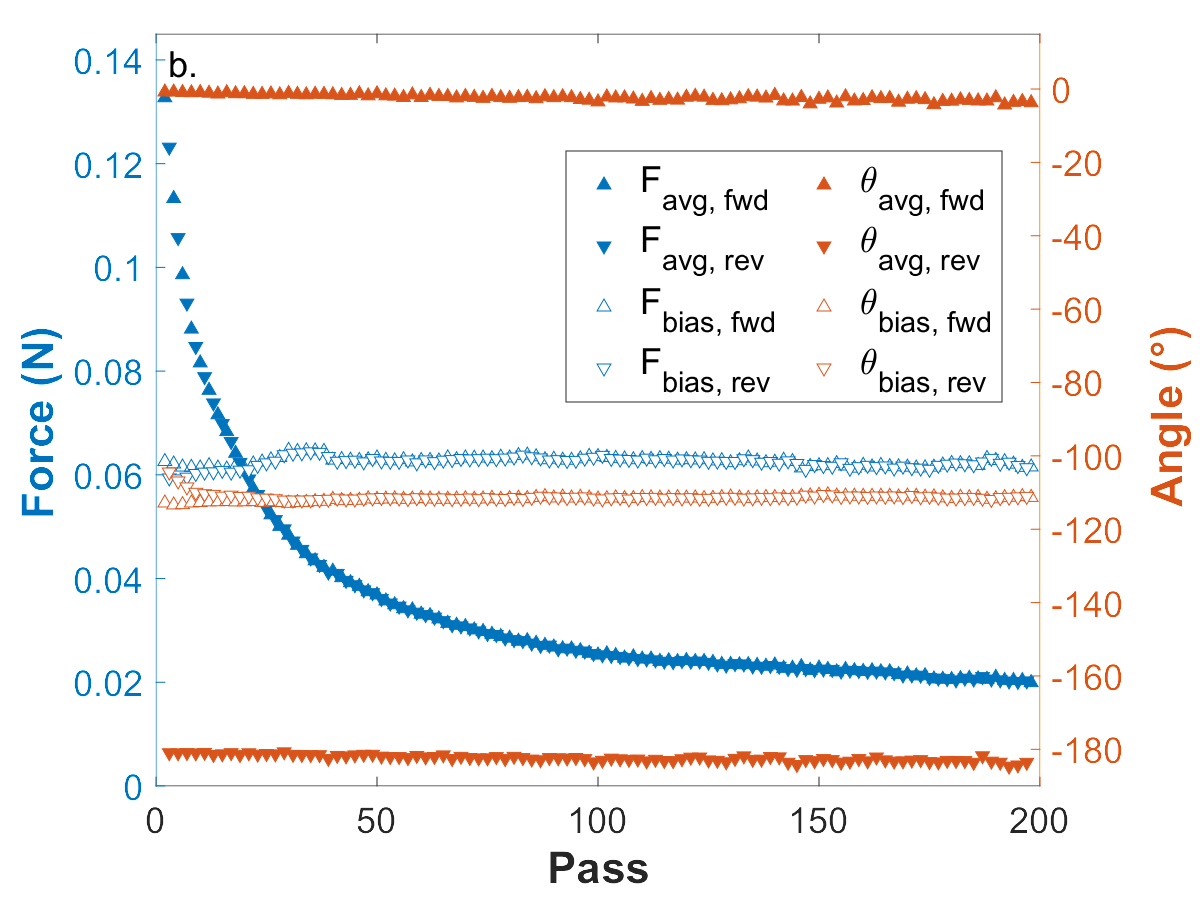}
	\caption{Evolution of the norm $F_{avg}$ (blue, filled symbols) and angle $\theta_{avg}$ (red, filled symbols) of the corrected horizontal force $\Vec{F}_{avg}$ and of the norm $F_{bias}$ (blue, open symbols) and angle $\theta_{bias}$ (red, open symbols) of the residual horizontal force $\Vec{F}_{bias}$, (a) vs. time during reciprocating passes 189 \& 190 and (b) vs. number of sliding passes for forward ($\blacktriangle$) and reverse ($\blacktriangledown$) motion.}
	\label{fig:UHV4Corr}
\end{figure}

This correction is applied to the data plotted on Fig.~\ref{fig:UHV2Raw}, and the results are plotted on Figure~\ref{fig:UHV4Corr}. During a reciprocating cycle (Fig.~\ref{fig:UHV4Corr}.a), which combines two consecutive passes in opposite sliding directions, the angle $\theta_{avg}$ of the average horizontal force $\overrightarrow{F_{avg}}$ is changing by $180\degree$ between forward and reverse direction, as expected for a reciprocating motion. On the contrary, the angle $\theta_{bias}$ of the bias force $\overrightarrow{F_{bias}}$ remains constant for both sliding directions (at around $-110\degree$): this force doesn't change with the sliding direction, and is indeed a bias force, which is not involved in the friction. Figure~\ref{fig:UHV4Corr}.b shows the evolution of the averaged values for each pass of the experiment. The angle $\theta_{avg}$ indeed oscillates between $-5\degree$ for forward passes and $-185\degree$ for reverse passes, while $\theta_{bias}$ remains almost constant (at around $-110\degree$), confirming the tendency observed on a single sliding cycle (Fig.~\ref{fig:UHV4Corr}.a).

The approach proposed here to combine measurements in forward and reverse direction is in fact similar to the one proposed by Burris \& Sawyer for low friction measurements\cite{burris_addressing_2009}, except that we apply it here with vectors instead of algebric values, thanks to the 6 axes force sensor instead of the traditional two sensor design of tribometers. The misalignment between the $z$ axes of the force sensor and of the upper manipulator is indeed a significant source of uncertainty, and a great advantage of our 6 axes force sensor is the ability to detect and compute such misalignment. The almost constant bias force $F_{bias}$ is about 0.06\,N here for a normal force $F_z$ of 3\,N, thus corresponding to a small misalignment of about $1.15\degree$.

\begin{figure}[hbt!]
	\centering
	\includegraphics[width=0.95\linewidth]{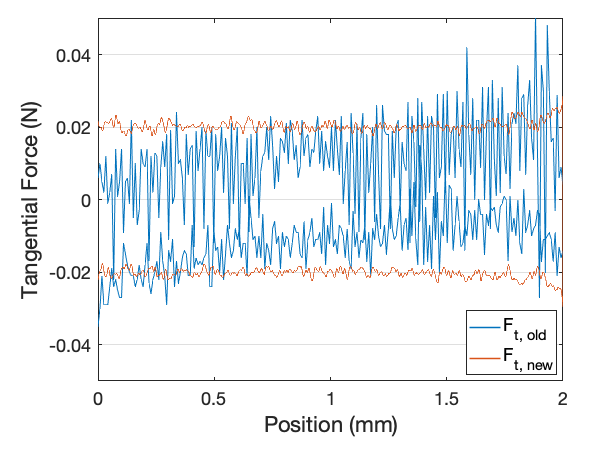}
	\caption{Evolution of tangential force as a function of position for sliding cycle 85 for two experiments, one performed on the old UHV tribometer of Ecole Centrale de Lyon~\cite{fontaine_JET222}, compared to the one presented in Figure~\ref{fig:UHV4Corr}.}
	\label{fig:UHV5Comp}
\end{figure}

In order to compare with the signals obtained using the traditional type of sensors used on most tribometers, Fig.~\ref{fig:UHV5Comp} compares the tangential force vs. position during one reciprocating cycle (cycle \#85), from the  experiment described above and from a similar older experiment (Fig.~4.a of reference~\cite{fontaine_JET222}) performed on the previous vacuum tribometer of Ecole Centrale de Lyon~\cite{martin_tribochemistry_1999,lemogne1999}. This previous tribometer was designed with force sensors outside of the vacuum chamber: a tangential force sensor (piezoelectric sensor) on an horizontal shaft with vertical motion (for application of a controlled normal load), and a normal force sensor (compliant plate with displacement sensor) on a vertical shaft with lateral motion (for linear reciprocating motion). For comparison purposes, on this figure, the tangential force is positive on the forward part of the cycle, and negative on the reverse part, thus delineating a tangential force loop. With the new 6 axes force sensor and its FPGA target (see section \ref{sec:FPGA}), the data are acquired at 2~kHz, allowing for a feedback loop to maintain the applied normal force constant, while with the old device only 256 data points were recorded along the track, corresponding to an acquisition rate of only 64~Hz, with no feedback loop on the normal force. Despite the higher acquisition rate, the signal to noise ratio is drastically improved, with the additional advantage that the new design is much stiffer than on the old tribometer.

To conclude this first example, we have shown that 6-axes force/torque sensors provide the unique opportunity to measure the unavoidable misalignment between the force sensor's axes and that of the actuation used to load the contact. Such a knowledge opens for specific signal analysis methods enabling elimination of the misalignment-induced force bias. Only similar de-biased signals can be safely used to characterize the friction properties of very low friction interfaces, like DLC/metal contacts under ultra-high vacuum.

To conclude this subsection on low friction measurements inside a vacuum chamber, we can list the following advantages with this 6 axes force/torque sensor over traditional tribometer designs:
\begin{itemize}
    \item sensor for force measurements and actuators for sample motion are handled separetely, allowing for a much stiffer design of the tribometer;
    \item all forces are measured on the same sensor, which allows the measurement of misalignments between the actuators and the sensor;
    \item thanks to implementation on a FPGA target, the forces can be measured at a high frequency (2~kHz so far), allowing for a continuous control of the applied force thanks to a feedback loop;
    \item the signal to noise ratio is significantly improved compared to previous design, allowing the precise measurement of very low friction coefficients.
\end{itemize}

\subsection{Elastomer friction}\label{sec:PDMS}

\begin{figure}[hbt!]
	\centering
	\includegraphics[width=0.99\columnwidth]{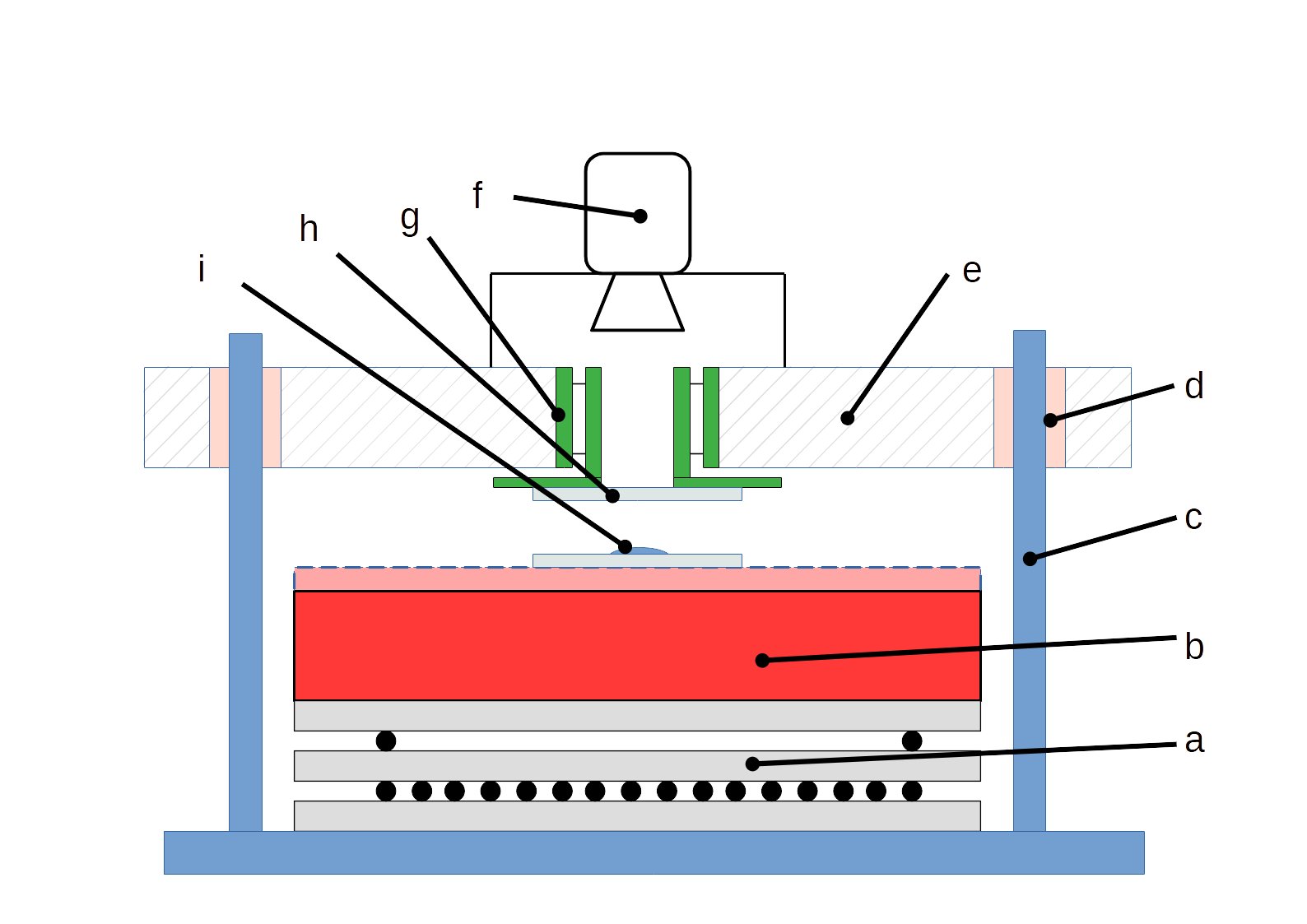}
	\caption{Sketch of the tribometer, where (a) is a $xy$ table, (b) is a $z$ piezoelectric actuator, (c) and (d) are precision rods and bushings for coarse vertical positioning of the top plate, (e) is the top plate holding both a high resolution camera (f) allowing observation of the contact area during the test and the six-axes force/torque sensor (g). The flat glass sample (h) is attached to the force sensor, while the spherical PDMS sample (i) is attached to the moving part of the machine, allowing the contact area to be observed while the $xy$ table moves.} \label{fig:mistress}
\end{figure}

In contrast with section~\ref{sec:DLC} where we considered a very-low-friction system, here we will perform measurements on a high-friction interface between a smooth elastomer sphere (PolyDiMethylSiloxane, PDMS, Sylgard 184, radius of curvature 9.42\,mm) and a smooth glass plate. The very same tribological pair has been extensively described in the literature~\cite{sahli_evolution_2018,mergel_continuum_2019,sahli_shear-induced_2019,papangelo_shear-induced_2019,lengiewicz_finite_2020}. Here it is tested in an apparatus sketched in Fig.~\ref{fig:mistress}, where the glass plate is directly attached to a single-stage version of our 6-axes force/torque sensor. The latter is made of stainless steel and its dimensions are the ones reported in Appendix~\ref{sec:appA}.

\begin{figure}[htb!]
	\begin{center}
	\includegraphics[width=0.8\columnwidth]{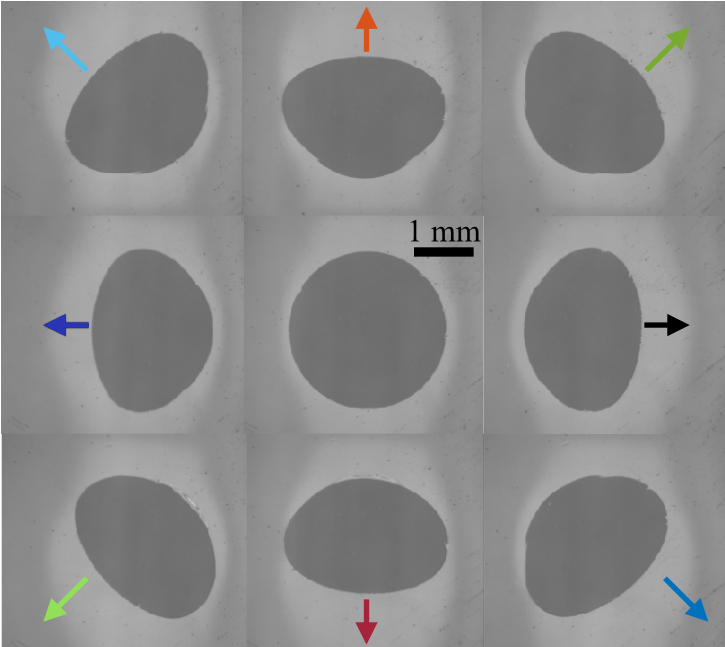}
	\caption{Contact area imaged at the beginning of the experiment (central picture) and during steady sliding (the eight peripheral pictures), along the eight directions used during the test. Arrows: direction of the motion of the PDMS sphere.} \label{fig:contacts}
	\end{center}
\end{figure}

The principle of the friction test is the following. A contact between PDMS and glass is first created by moving upward the piezoelectric actuator, until a desired vertical load $F_{z_{0}}=0.5$\,N is reached. The altitude of the PDMS sphere is then adjusted in real time through a FPGA-controlled feedback loop in order to keep the vertical load constant during the rest of the test. Shear is then imposed to the contact by moving the $xy$ table with a 1\,mm stroke, along various directions in the $xy$-plane (see Fig.~\ref{fig:contacts}), at a constant velocity of 0.1\,mm/s. Note that before the experiment, the alignment of the glass substrate (attached to the 6-axes sensor) with the plane of motion of the $xy$ table has been realized within a few $\mu$m/mm in each direction. During all the test, the contact area is filmed with a resolution of 2000 x 3008 pixels at a frame rate of 4 images per second. Indeed, the specific hollow design of our force/torque sensor, together with the transparency of the contacting materials (glass and PDMS), make it possible to perform in situ, in operando observations of the contact interface (see Fig.~\ref{fig:mistress}).

Figure~\ref{fig:contacts} shows various typical contact images taken during the experiments, through the central hole of the force/torque sensor. The central image corresponds to the initial contact, when the PDMS sphere is simply pressed onto the glass plate, i.e. with no shear loading. The other eight images are the final images, i.e. in steady sliding regime, in all the eight directions of motion tested.
As one can see, in all cases, the shape and area of the contact have changed with respect to the initial contact, in good agreement with previous results from the literature~\cite{sahli_evolution_2018,mergel_continuum_2019,papangelo_shear-induced_2019,lengiewicz_finite_2020}: while the initial contact is perfectly circular, the steady-state contact is more ellipse-like. One can clearly see here that the ellipse's major axis is always orthogonal to the motion direction, but otherwise the contact area and shape are the same for all eight experiments.

\begin{figure}[htb!]
	\begin{center}
    \includegraphics[width=0.99\columnwidth]{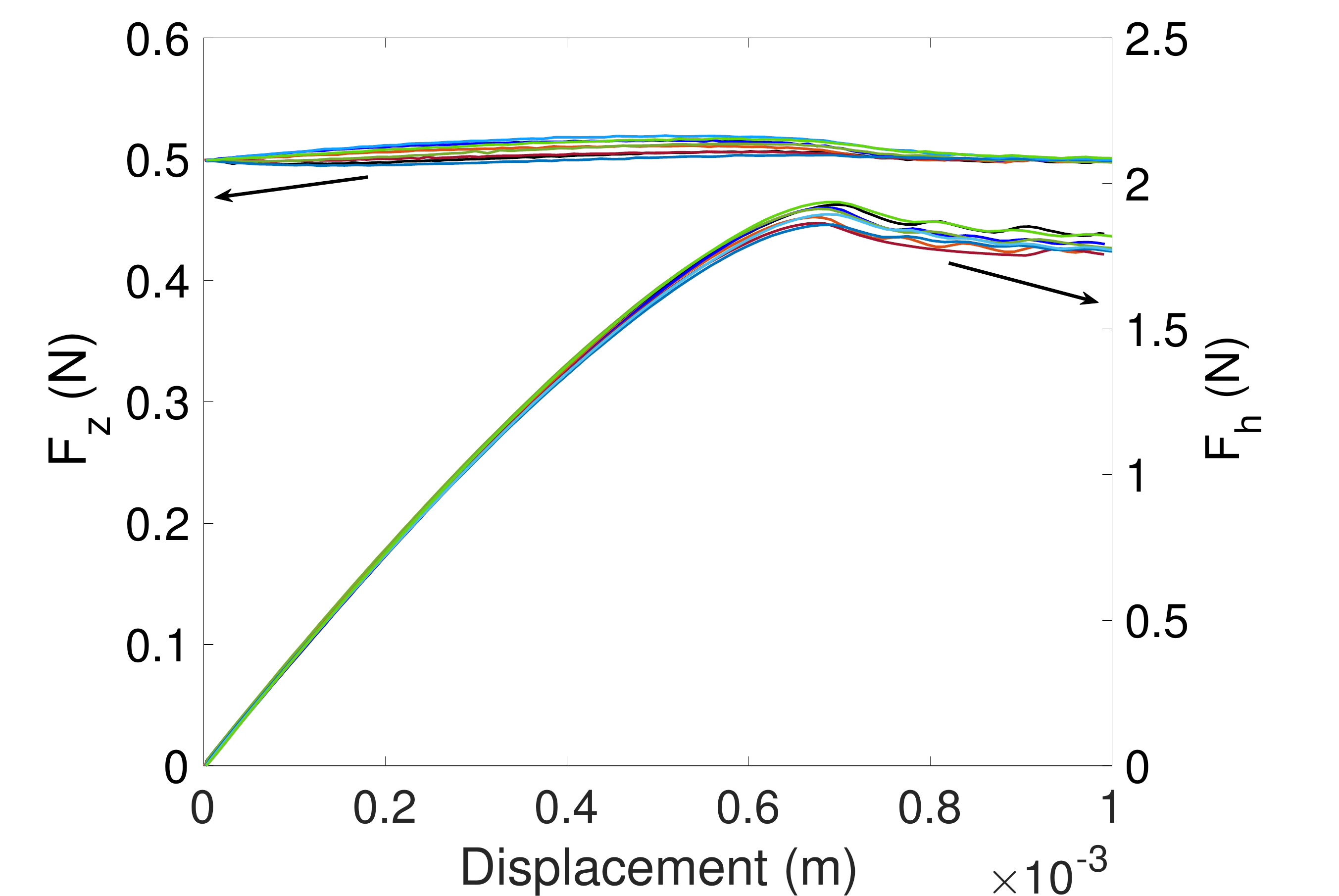}
	\caption{Contact forces as a function of tangential displacement (in the $xy$ plane): Normal, $F_z$ (left axis) and tangential, $F_{h}$ (right axis), for the eight directions used during the tests. The color code corresponds to that of the arrows in Fig.~\ref{fig:contacts}. \label{fig:z_defaut}}
	\end{center}
\end{figure}

Figure~\ref{fig:z_defaut} shows the evolution of the three components of the contact force: the normal force, $F_z$ (left axis), and the tangential force, $F_{h}=\sqrt{F_x^2+F_y^2}$ (right axis), as a function of the displacement in the $xy$ plane, for all eight runs. The normal force is found essentially constant all along the experiments, as expected in the controlled-loop conditions used. Concerning the tangential force, all eight curves do overlap nicely, showing that our sensor indeed enables robust measurements in all directions of motion. For all experiments, the tangential load is initially zero, then increases as the contact is shear-loaded, reaches a maximum corresponding to the static friction peak, and only then enters a rather steady macroscopic sliding regime. The ratio of the peak force to the corresponding contact area ($F_{h, peak}/A_{peak}$), is found very reproducible with respect to the sliding direction, with a value of $0.44 \pm 0.02$\,MPa (the error bar is the standard deviation over the eight directions and over three repetitions for each direction). Both the above-mentioned qualitative and quantitative features are in very good agreement with the literature~\cite{sahli_evolution_2018,mergel_continuum_2019,lengiewicz_finite_2020}, which confirms the applicability of our 6-axes sensor to such high-friction tribological situations. The multi-directionality of the measurements now opens possibilities to explore much richer contact kinematics than the usual linear stroke friction experiments, while keeping the possibility to optically monitor the contact area simultaneously.

\subsection{Contact localization}\label{sec:localization}
The previous two examples have illustrated the possibility offered by our 6- axes sensor to access the three components of contact forces, but the torque values were not explicitly taken advantage of. Here, we will make use of all six outputs of the sensor to monitor the evolution of the point of application of forces in a moving contact. We use the same tribometer as in section~\ref{sec:PDMS} (Fig.~\ref{fig:mistress}), but we replace the PDMS sphere with a steel ball of diameter 12.7\,mm (to reduce the contact area) and insert a droplet of oil between steel and glass (to reduce the friction force). We perform the same eight runs in the same eight directions as in section~\ref{sec:PDMS}, under a constant (closed-loop controlled) normal load of 1\,N, and at a constant driving velocity of 0.1 mm/s. In those conditions the friction force is about 0.135\,N.

When the contact zone is infinitely small, then the wrench of contact forces reduces to a single force applied at point $H$. Ideally, the moment at $H$ is zero and therefore $H$ belongs to the screw axis. When the contact zone remains small, we may introduce the center of contact as the point $H$ where the moment is minimal. This point belongs again to the screw axis.

Let $\overrightarrow{R}=(F_x, F_y, F_z)$ and $\overrightarrow{M}=(M_x, M_y, M_z)$ be the force and torque vectors at point $O$ defined by the intersection between the sensor axis (as defined from its calibration) and the plane $z=0$. This pair of vectors form a screw $\left\{\overrightarrow{R}, \overrightarrow{M}\right\}_O$.

The screw axis has direction $\overrightarrow{R}$ and passes through the point $A$ such that:
\begin{equation}
    \overrightarrow{OA}=\frac{1}{R^2}\cdot\overrightarrow{R}\times \overrightarrow{M}.
\end{equation}
For any other point $H$ on the axis, it exists $\lambda$ such that, 
\begin{equation}
    \overrightarrow{OH}=\overrightarrow{OA}+\lambda\cdot\overrightarrow{R}.
\end{equation}
Looking for $H$ in the plane $z=0$, we have
 $\overrightarrow{OH}\cdot\overrightarrow{z}=0$, and therefore
\begin{equation}
    \lambda=-\frac{\overrightarrow{OA}\cdot\overrightarrow{z}}{\overrightarrow{R} \cdot \overrightarrow{z}}
\end{equation}
which leads to:
\begin{equation}
    \overrightarrow{OH}=\frac{1}{R^2}\cdot\overrightarrow{R}\times\overrightarrow{M}-\frac{\overrightarrow{OA}\cdot\overrightarrow{z}}{\overrightarrow{R}\cdot\overrightarrow{z}}\cdot\overrightarrow{R}.
\end{equation}
The projection of this equation along both the $x$ and $y$ axes ($z$ being zero by assumption) leads to:
\begin{equation}
    \begin{pmatrix} x\\y \end{pmatrix}=\frac{1}{R^2}\cdot\begin{pmatrix}
    F_y\cdot M_z-F_z\cdot M_y -(F_x\cdot M_y-F_y\cdot M_x)\cdot \frac{F_x}{F_z} \\F_z\cdot M_x-F_x\cdot M_z-(F_x\cdot M_y-F_y\cdot M_x)\cdot\frac{F_y}{F_z} \end{pmatrix}\label{eq:positions}
\end{equation}

\begin{figure}[hbt!]
	\centering
	\includegraphics[width=0.9\linewidth]{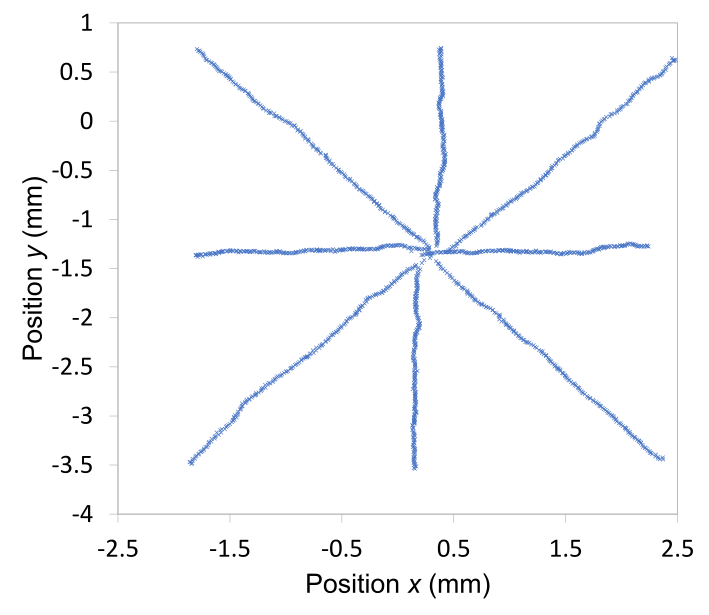}
	\caption{Coordinates $x$ and $y$ of all contact points during the eight runs, as obtained from the six measured components of the forces/torques (Eq.~\ref{eq:positions}). $(x=0,y=0)$ corresponds to the center of the force sensor.}
	\label{fig:Pos_from_forces}
\end{figure}

These coordinates are plotted in Fig.~\ref{fig:Pos_from_forces} for each measurement points of the eight successive tests. Note that point $O$ has coordinates $x=0$, $y=0$. As  can be seen, our evaluation successfully allows to draw the trajectories of the contact during the eight runs. In particular, the directions and length of the eight motions are well identified. We emphasize that such a localization procedure can be performed only because our sensor provides all six components of the forces/torques due to the contact. Such localization may be an asset when, for some reason, a position sensor cannot be installed close to the contact, as is often the case when working in Ultra High Vacuum for instance. Note that the facts that the trajectories are not perfect lines and that all trajectories do not start exactly at the same point are not defects of our driving apparatus, but rather reveal some imperfections in the inversion, such as (i) noise in the force/torque signals, (ii) contact forces not being applied to a single point, but rather distributed over a finite-sized contact zone, and (iii) actual contact altitudes being slightly different form the assumed $z=0$.

\section{Conclusions}
We have described the general design of a novel flexure-based 6-axes force/torque sensor which avoids any backlash or orthogonality imperfection between force components, and is monitored non-invasively via non-contact displacement sensors. The stiffness associated with the various force/torque components can be tuned independently, and the sensor's hollow structure enables visualization or further system instrumentation through its central axis. We have realized and calibrated two such 6-axes sensors, and mounted them on two different tribometers. 

To validate the capabilities of our sensors, we have performed three types of experiments, on both very-low and very-high friction interfaces, in either ambient or ultra-high-vacuum conditions, and with or without in situ contact visualization. Those applications required the measurement of at least all force components (for friction coefficient/stress evaluations along different sliding directions), and even of all six force/torque components (for contact localization without the help of any additional displacement sensor).

Although our sensor was initially motivated by, and used in, tribological contexts, we expect that our new design can be useful in any other type of mechanics application where the six components of the forces/torques are of interest, and/or when specific additional constraints apply, including sample visualization or severe environmental conditions. It can be noted that direct access to the area of interest, thanks to the hollowed structure, makes it possible to consider instrumenting the device relatively simply with other measurement techniques, including Infra-red or Raman spectroscopy and laser vibrometry.

\section*{author's contributions}\label{author's contributions}
M.G and T.D. designed the sensor. J.S., D.D., J.F. and A.L.B designed the research. M.G., T.L.M, J.F. and C.O. performed the research. M.G., J.S., D.D., A.L.B. and J.F. wrote the manuscript. All authors commented on the manuscript.

\section*{acknowledgments}
This work was supported by the Agence Nationale de la Recherche under grant No. ANR-11-NS09-01. It was also supported by the LABEX iMUST (ANR-10-LABX-0064) of Université de Lyon, within the program "Investissements d'Avenir" (ANR-11-IDEX-0007) operated by the French National Research Agency (ANR).

We thank PREVAC company (\url{https://www.prevac.eu}) for useful discussions about the integration of the sensor in UHV chamber.  

\section*{data availability}
The data that support the findings of this study are available from the corresponding author upon reasonable request.


%

\appendix

\section{Dimensions of the second sensor}\label{sec:appA}

The sensor used in the experiments of sections~\ref{sec:PDMS} and \ref{sec:localization} consists of a single stage: the first stage of section \ref{sec:gendesign}. Its dimensions are $R=140\:\text{mm}$, $r=112\:\text{mm}$, $H=18\:\text{mm}$, $L=4\:\text{mm}$, $l=6\:\text{mm}$, $e=1\:\text{mm}$.

Its calibration matrix is:
\begin{widetext}

\begin{equation}
A=
\begin{pmatrix*}[r]
14.467	&  -23.144 &   12.307 &   -0.380 &  -26.977 &   23.476 \\ 
-20.684 &   12.473 &   22.301 &  -24.968 &   -2.088 &   13.351\\
 -1.051 &   -1.317 &   -3.572 &   -0.565 &   -0.147 &   -4.013\\
584.104 & -267.679 & -508.896 &  617.732 &  -65.546 & -409.340\\
301.457 & -654.779 &  421.871 &   78.498 & -699.030 &  563.909\\
348.506 & -343.401 &  329.638 & -331.954 &  357.993 & -357.581\\
\end{pmatrix*} 
\label{matrice_passage}
\end{equation}
\end{widetext}

\end{document}